\documentclass[12pt]{iopart}

\usepackage{iopams}
\usepackage{setstack}
\usepackage{graphicx}
\usepackage{cite}

\newcommand{\degree}{\ensuremath{^\circ}}%
\newcommand{\cost}{\ensuremath{\langle\cos^2\theta_\mathrm{2d}\rangle}}

\newlength{\figwidth}
\newlength{\figwidthsmall}
\begin{document}

\title[Orientation-dependent ionization yields from fixed-in-space molecules]{Orientation-dependent
  ionization yields from strong-field ionization of fixed-in-space linear and asymmetric top
  molecules}

\author{Jonas L. Hansen$^1$ , Lotte Holmegaard$^2$ , Jens H. Nielsen$^3$ and Henrik Stapelfeldt$^{1,2}$}%
\address{$^1$\,Interdisciplinary Nanoscience Center (iNANO)\\
    $^2$\,Department of Chemistry, Aarhus University, 8000 Aarhus C, Denmark \\
  $^3$\,Department of Physics and Astronomy, Aarhus University, 8000 Aarhus C,
   Denmark}

\author{Darko Dimitrovski and Lars Bojer Madsen}
\address{Lundbeck Foundation Theoretical Center for Quantum
System Research, Department of Physics and Astronomy,
Aarhus University, 8000 Aarhus C, Denmark}


\begin{abstract}
  The yield of strong-field ionization, by a linearly polarized probe pulse, is studied
  experimentally and theoretically, as a function of the relative orientation between the laser
  field and the molecule. Experimentally, carbonyl sulfide, benzonitrile and naphthalene
  molecules are aligned in one or three dimensions before being singly ionized by a 30 fs
  laser pulse centered at 800 nm. Theoretically, we address the behaviour of these three molecules. 
  We consider the degree of alignment and orientation and 
  model the angular dependence 
  of the total ionization yield by molecular tunneling theory accounting for the Stark shift of the 
  energy level of the ionizing orbital.
  For naphthalene and benzonitrile the orientational dependence of
  the ionization yield agrees well with the calculated results, in particular the observation that
  ionization is maximized when the probe laser is polarized along the most polarizable axis. For OCS
  the observation of maximum ionization yield when the probe is perpendicular to the internuclear
  axis contrasts the theoretical results. 
  \end{abstract}


\maketitle

\section{Introduction}

The ionization step leading to single ionization in the multiphoton or tunnel ionization regime is a fundamental process and a first step that triggers the subsequent dynamics in attoscience \cite{Corkum1993}, both in the process of high-order harmonic generation and in streaking experiments \cite{AttoReview}. This step is thought to be well understood for atoms, however for larger molecules much less is known. Of particular importance is the understanding of the dependence of the initial ionization step on the molecular orientation with respect to the external field. Commonly used theories to explain strong-field phenomena extended to larger systems are the molecular tunneling theory (MO-ADK) \cite{tong:02} and the strong-field approximation (SFA) \cite{Faisal:2000,kjeldsen:04}. Recent experiments, however, have shown disagreement with these theories \cite{pavicic:07}, and the interpretation of the underlying physics has been and still is an area of debate \cite{Mahmoud:pra:2009,CDLinPRA2009,SaenzPRL2010,WuPRA2011,GallupPRA2010,Murray:PRL:2011,SpannerPRA2009}. To fully test existing theories and to guide the way for new theory development, experiments extended to larger and more complex molecular systems are needed.

Experimental orientation-dependent ionization rates have only been obtained for small molecular systems such as N$_2$, O$_2$, CO, CO$_2$ and C$_2$H$_2$ \cite{litvinyuk:03,alnaser:04,alnaser:05,pavicic:07}. Many of these previous experiments have involved several steps when extracting orientation-dependent ionization rates from the measurements which complicates the interpretation. The direct measurement of the angular dependence of ionization was reported in \cite{pavicic:07}, where the molecules were aligned non-adiabatically, demonstrating the essential need for creation of an ensemble of highly aligned molecules. An alternative to pre-aligning the molecules is to infer their spatial orientation after the ionization event. In favorable cases, this can be accomplished if the ionization process leads to fragmentation of the molecule and the recoil directions of the ionic fragments uniquely defines the molecular orientation at the time of ionization. Recently,  this approach was used to determine the orientation-dependent ionization yields for HCl molecules ionized by intense circularly polarized infrared femtosecond laser pulses \cite{akagi:science:2009}. For larger molecular species it is mostly impossible to access molecular frame information using fragmentation measurements.  This fact is very important 
 since it puts more than technical constraints, like low counting statistics, on the type of systems that can be investigated by recoil methods.
 In addition,  the often encountered situation where ionization does not cause fragmentation, also can not give access to orientation-dependent measurements by recoil spectroscopy. 

Hence, in general, molecular frame measurements require pre-alignment or pre-orientation of the molecules. The latter approach was employed recently in a series of joint experimental and theoretical studies on molecular frame photoelectron angular distributions (MFPADs) for three different adiabatically aligned and oriented molecules, carbonyl sulfide (OCS), benzonitrile (C$_7$H$_5$N, to be denoted by BN in the following) and naphthalene (C$_{10}$H$_8$, to be denoted by NPTH in the following) molecules, ionized by intense infrared {\it circularly} polarized femtosecond pulses \cite{Holmegaard10,Hansen:PRA:2011,Dimitrovski:PRA:2011,Hansen:PRL:2011}. The theoretical description of the MFPADs was accomplished by amending the existing theories of strong-field ionization to account for the Stark shift induced by the molecular dipole and polarizability
\cite{Holmegaard10,Dimitrovski:PRA:2010,Dimitrovski:PRA:2011}. Here, we employ {\it linearly} polarized pulses to perform measurements on the orientational dependence of the ionization yield on OCS, BN and NPTH molecules and use the same theories to describe the experimental results. 

The paper is organized as follows. In the next section we describe the experimental setup. Section 3, organized into subsections 
corresponding to the different molecular targets, contains detailed comparison of the experiment and theory. Section 4 concludes.

\section{Experimental Setup}
The experimental setup was recently described in detail \cite{Hansen:PRA:2011} so the discussion here
is brief. A mixture of 90 bar helium and either 10 mbar OCS, 5 mbar BN or 5 mbar NPTH is expanded into vacuum through a pulsed valve. The resulting molecular beam
is collimated by two skimmers before entering a 15 cm long electrostatic deflector. Traversing the
deflector, polar molecules (OCS, BN) are dispersed along the gradient of the field according to
their effective dipole moment which depends on the individual rotational quantum states of the
molecules \cite{filsinger_quantum-state_2009, holmegaard_laser-induced_2009}. This dispersion 
allows for quantum state selection of the molecular ensemble in the interaction region downstream
from the deflector where the molecular beam is crossed by one or more laser beams producing either
adiabatically or non-adiabatically aligned targets. Alignment in the adiabatic regime is achieved
using the output from an injection seeded Nd:YAG laser (20 Hz, $\tau_{\mathrm{FWHM}}$ = 10 ns, $\lambda$ =
1064 nm), whereas field-free alignment, only implemented for OCS, is obtained using a part of the
output from a 1 kHz amplified Ti:sapphire laser (800 nm) stretched to 0.5 ps and performing the
experiment at the peak of the half revival 40.5 ps after the non-adiabatic alignment pulse.  The degree of
alignment is initially measured by Coulomb exploding the molecules using an intense femtosecond
probe laser pulse (1 kHz, 30 fs FWHM, 800 nm) and detecting the axially recoiling S$^+$ fragments from
OCS and CN$^+$ fragments from BN which is described elsewhere \cite{Holmegaard10}. Unlike OCS and
BN, NPTH does not contain good observables which allow for a precise measurement of the degree of
alignment. Confinement of the molecular axes  can, however, still be demonstrated through H$^+$ ion
imaging, providing information about the spatial direction of the molecular plane~\cite{Hansen:PRL:2011}.

Using similar alignment laser parameters but inverting the voltage on the velocity-map-imaging (VMI)
spectrometer, electrons rather than ions are focused onto the detector. In the experiment the
(major) polarization axis of the alignment pulse is kept fixed along the static field axis of the VMI spectrometer whereas the angle, $\Theta$, of the linearly polarized
probe pulse with respect to the static field is varied between 0$^\circ$ and 180$^\circ$.
Using this particular alignment geometry the direction of the dipole moment for OCS and BN will be parallel to the static field breaking the head-for-tails symmetry of the ensemble to produce oriented molecules.
The degree of orientation is estimated to $\sim80\%$ for both OCS and BN \cite{Holmegaard10}. In this experiment the alignment dependent ionization yield will, however, not hold any information on the orientation, since the probe pulse contains more than a few cycles and 
has an uncontrolled carrier-envelope phase.

In all measurements the intensity of the probe pulse has been adjusted to a regime where the molecules investigated primarily undergo single ionization without
fragmentation. 
As a consequence the electrons produced will be from the ionization step alone and it
is, therefore, possible to obtain orientation-dependent ionization yields by integrating the
measured electron signal at the various angles between the probe and alignment laser. To prevent
fluctuations in the laser intensity from affecting the measurements, the total ionization yield has been
normalized to that obtained from unaligned molecules in all of the experiments.

\section{Experimental results and comparison with theory}

\subsection{Carbonyl sulfide, OCS}
\begin{figure}
    \centering
    \includegraphics[width=0.65\textwidth]{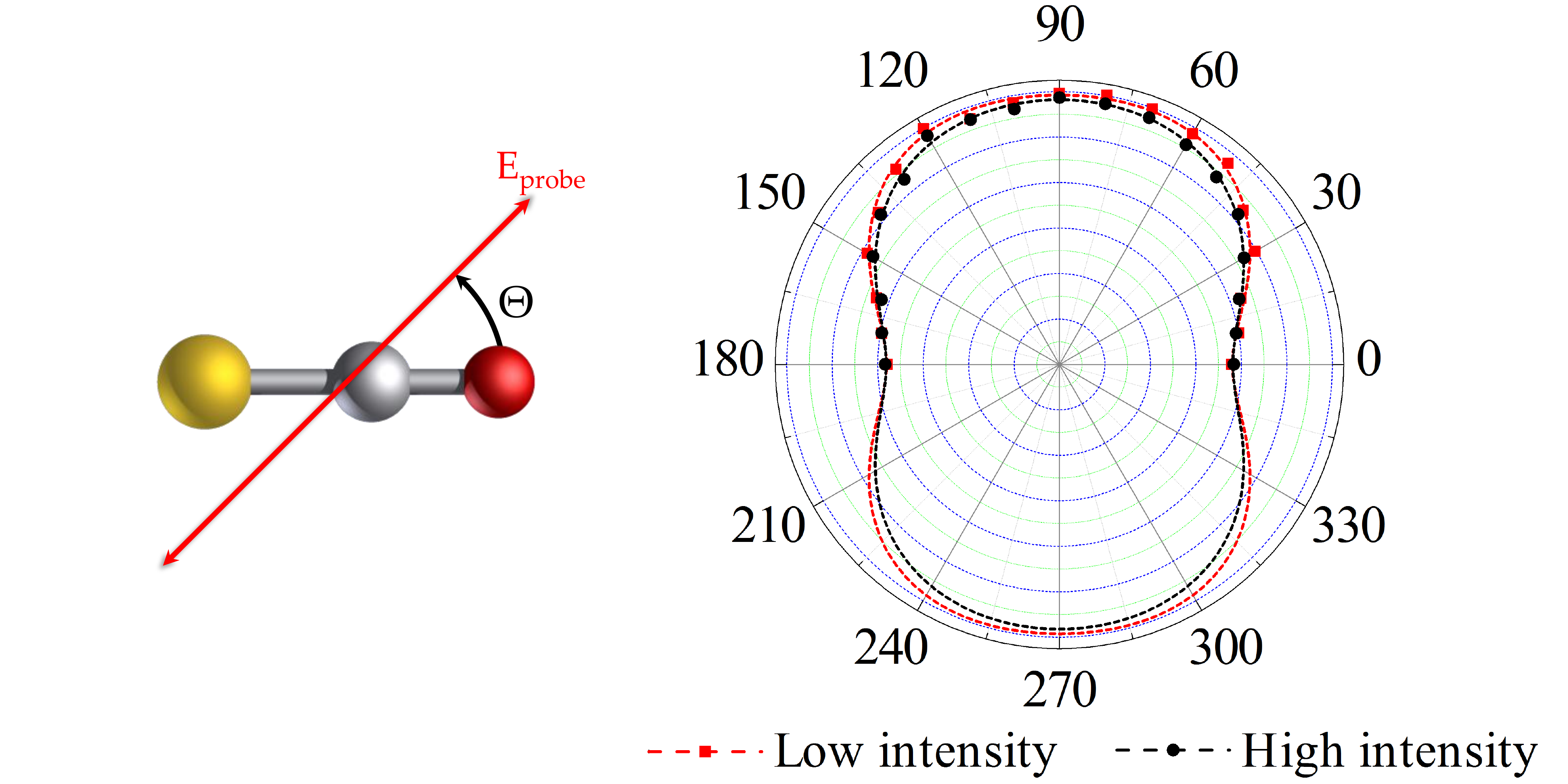}
    \caption{
Total Ionization yield of adiabatically aligned OCS molecules as a function of the angle between the polarization axes of the alignment and probe pulses. The ionization yield is measured at two different intensities of the probe beam; $\mathrm{I}_{\mathrm{probe, high}}~=~1.8\times10^{14}$~Wcm$^{-2}$ and $\mathrm{I}_{\mathrm{probe, low}}~=~1.5\times10^{14}$~Wcm$^{-2}$ (30 fs, 800 nm). The filled squares and circles represent the measurements and the dashed curves, to guide the eye, are Legendre polynomial fits.}
    \label{OCS_ad}
\end{figure}

In order to measure the angular dependence of the total ionization yield a high degree of alignment is
required. This is achieved by inducing adiabatic alignment of OCS molecules prepared in the
lowest rotational quantum states \cite{Nielsen:PCCP:2011}. Using a linearly polarized alignment pulse, polarized in the plane of the
detector, with an intensity of $\mathrm{I}_{\mathrm{YAG}}~=~8.4\times10^{11}$~Wcm$^{-2}$
(10 ns, 1064 nm) a degree of alignment of \cost~$=0.90$ is obtained, $\theta_{\mathrm{2d}}$ being
the angle between the YAG polarization and the molecular axis. 
Figure~\ref{OCS_ad} displays the ionization yield, obtained by measuring the total electron signal, as a function of 
the angle between the most polarizable axis and the linear polarization, 
$\Theta$, for two different intensities of the probe pulse.
Figure~\ref{OCS_ad} shows that there is hardly any difference between the orientational
dependence of high and low probe intensities, both showing a clear minimum of the signal when
$\Theta=0$\degree, and a maximum in the ionization yield at $\Theta=90$\degree. The two
degenerate highest occupied molecular orbitals (HOMOs)  of OCS have nodal surfaces coinciding with the internuclear axis.
For a highly aligned ensemble of OCS molecules there will be no or little contribution from the orbital density of the HOMOs which can participate in the ionization process when the field is strictly parallel to the internuclear axis. Some suppression of the emission along this axis is therefore expected. However, such an argument does not explain why there is a maximum in the ionization yield at $\Theta=90$\degree.

\begin{figure}
    \centering
\includegraphics[width=0.55\textwidth]{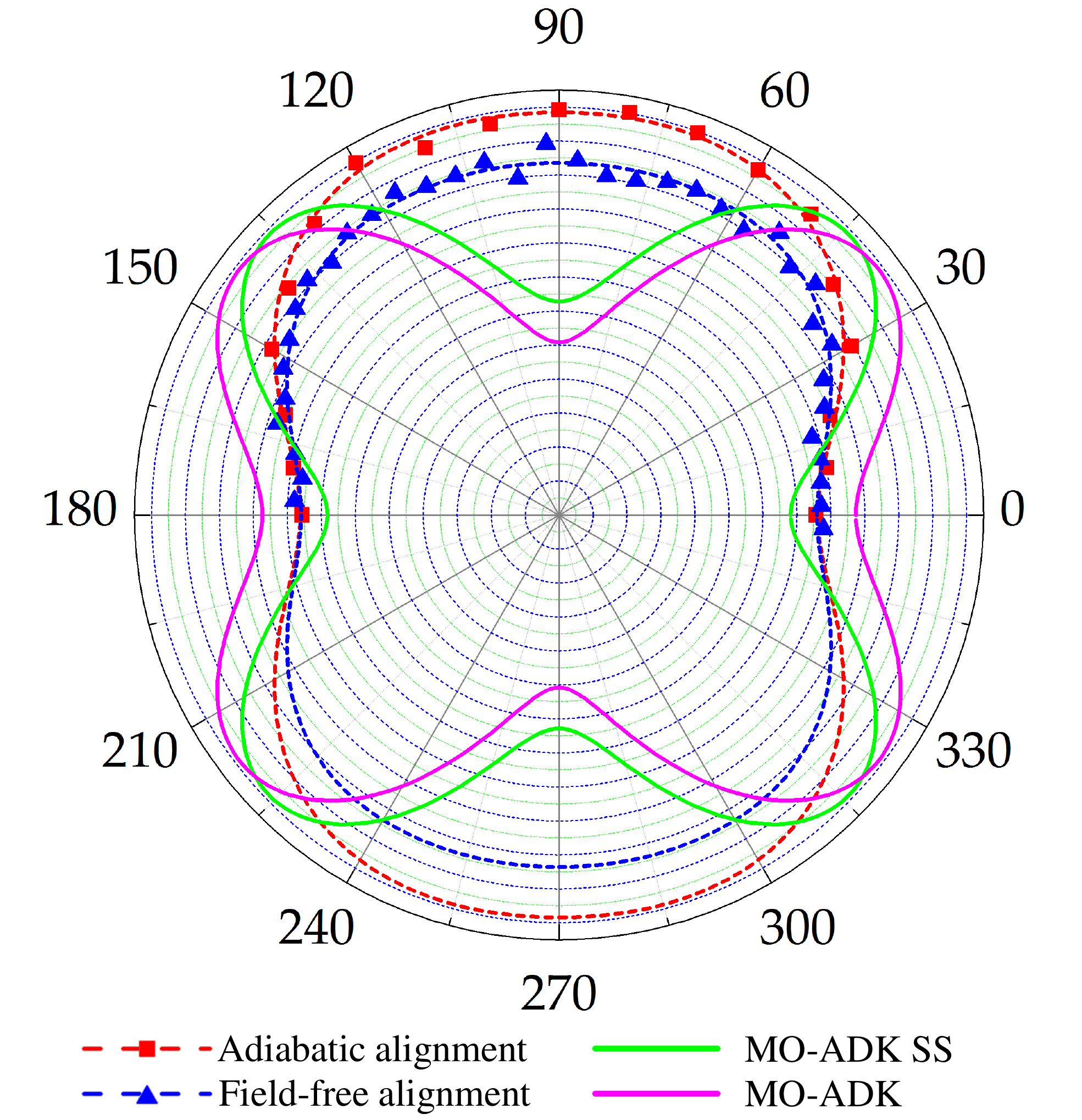}
\caption{
Total ionization yield of field-free aligned OCS (blue triangles, \cost~$=0.84$) and adiabatically aligned OCS molecules (red squares, \cost~$=0.90$) as a function of the angle between the polarization axes of the alignment and probe pulses. The filled symbols represent the measurements and the dashed curves, to guide the eye, are Legendre polynomial fits. The ionization yield is modeled using standard MO-ADK~\cite{tong:02} (solid magenta curve) and Stark-shift modified MO-ADK~\cite{Holmegaard10,Dimitrovski:PRA:2010,Dimitrovski:PRA:2011}(solid green curve). Both calculations assume a \cost~$=0.9$. The intensities of the laser pulses used for the experiments and theoretical calculations are $\mathrm{I}_{\mathrm{non-ad.}} = 8.3\times10^{12}$~Wcm$^{-2}$(0.5 ps, 800 nm), $\mathrm{I}_{\mathrm{YAG}}=7.7\times10^{11}$~Wcm$^{-2}$ (10 ns, 1064 nm) and $\mathrm{I}_{\mathrm{probe}}~=~1.5\times10^{14}$~Wcm$^{-2}$ (30 fs, 800 nm).

}
    \label{OCS_comparison}
\end{figure}

A reasonable concern of the experiment is if the presence of the YAG laser field
perturbs the ionization pathway and, in particular, the orientational dependence measured. To experimentally test this, the
orientational dependence of the ionization yields was also measured from field-free aligned OCS
molecules. Using similar experimental conditions, the measurements were repeated, replacing the
adiabatic pulse with a 0.5-ps-long non-adiabatic alignment pulse of intensity 
$\mathrm{I}_{\mathrm{non-ad.}}~=~8.3\times10^{12}$~Wcm$^{-2}$ (800 nm). At the peak of the half
revival of OCS, occurring $40.5$~ps after the alignment pulse, an alignment degree of \cost~$=0.84$ was obtained. The ionization yield was
extracted from the total electron signal measured at an intensity of the probe beam corresponding to
$\mathrm{I}_{\mathrm{probe, low}}~=~1.5\times10^{14}$~Wcm$^{-2}$ (30 fs, 800 nm).  Figure
\ref{OCS_comparison} provides a comparison of the orientation-dependent ionization yield of single
ionization of field-free aligned and adiabatically aligned OCS molecules. The ionization signal from
field-free aligned molecules exhibits an orientation dependence similar to that observed for
adiabatically aligned molecules, i.e., a clear suppression at $\Theta=0$ and a clear enhancement at
$90$\degree. The only discrepancy lies in the ratio $(R)$ between the yield, $Y$, at $\Theta=0$\degree\ and at
$90$\degree , 
\begin{equation}
R=\frac{Y(\Theta=0\degree)}{Y(\Theta=90\degree)}.
\end{equation}
For adiabatically aligned
molecules $R=0.65$ and for field-free aligned molecules $R=0.74$. We ascribe this difference
to the fact that the degree of alignment is stronger in the adiabatic case.

The permanent dipole moment of OCS results in large Stark shifts due to the probe pulse that should be considered 
when calculating the ionization yields \cite{Holmegaard10, Dimitrovski:PRA:2011}. Also
the alignment distribution of the
molecules should be taken into account. We use a Gaussian approximation for the alignment
distribution that can be used in cases of adiabatic alignment
\cite{friedrich_alignment_1995}. This alignment distribution is used
for all molecules analyzed here, the only input being $\cost$. Calculations are performed using the Stark-shifted corrected
\cite{Dimitrovski:PRA:2011} molecular orbital tunneling theory (MO-ADK) \cite{tong:02} and taking into account the HOMO, the dipole and the polarizability  of OCS, see \cite{Dimitrovski:PRA:2010}. The ionization potential of the HOMO is $I_p=11.4 $ eV and for the probe intensity in the experiment with adiabatic alignment the Keldysh parameter \cite{keldysh:65}  $\gamma \simeq 0.7$.

From figure \ref{OCS_comparison}, it is apparent that the
theory cannot reproduce the maximum for the ionization yields perpendicular to the molecular
axis. Instead, the ionization yields peak at approximately $45\degree$ with respect to the molecular
axis. Compared to the conventional MO-ADK theory \cite{tong:02}, the Stark-shifted MO-ADK theory represents an
improvement of around $10\degree$ when comparing to the experiment, see figure \ref{OCS_comparison}.

Previously, ionization of OCS with circularly polarized fields have been studied~\cite{Holmegaard10, Dimitrovski:PRA:2011}, and it was found that ionization occurs efficiently when the field points along the permanent dipole moment, i.e. along the molecular axis. In this connection, we note two important differences between strong field ionization induced by circularly polarized (CP) and by linearly polarized (LP) fields.

First, there is a fundamental difference relating to recombination and rescattering of the ejected electron. In CP fields, the wave packet of the tunneled electron does not revisit the parent ion, recombination and rescattering is minimized and the instantaneous ionization rate is mapped to the final momentum distributions. In LP fields this one-to-one mapping is lost. Our results point to the fact that the total ionization yields by LP laser pulses should not be calculated solely based on ionization rates, although such calculations have reproduced the experimental results for small molecules like ${\rm O}_2$ and ${\rm N}_2$ \cite{pavicic:07}.

Second, there is a difference relating to the influence of electronically excited states on the ionization process. For example in \cite{Kumarappan:PRL:2008}
comparison between theory and experiment on strong-field ionization of laser-aligned CS$_2$ indicated that excited states play a role.
For an arbitrary orientation between the external field and the molecular axis, the molecular 
orbital is no longer an $L_Z$ eigenstate in the laboratory fixed frame. 
Circular polarization, however, still changes the $\Lambda$ quantum number monotonously towards higher values. For linear polarization lower $\Lambda$ values will be favoured due to selection rules. 
The fact that within the excited state spectrum only states with relatively low $\Lambda$ are within reach by few-photon absorption means that excited states are expected to play a larger role for linear polarization than for circular polarization.
Along these lines,  we also note that the discrepancy between the experiment and the theory, observed for OCS, is similar to recent studies on ${\rm CO}_2$ \cite{pavicic:07}. In the case of ${\rm CO}_2$ different theoretical methods and possible explanations for the discrepancy included the influence of excited states \cite{Mahmoud:pra:2009}, the influence of HOMO-1 and HOMO-2 \cite{WuPRA2011}, and orbital modification and interference \cite{Murray:PRL:2011}. The debate is ongoing, see also Refs. \cite{CDLinPRA2009,SpannerPRA2009,SaenzPRL2010,GallupPRA2010}. In the case of OCS, all  candidates
proposed in the theories for ${\rm CO}_2$, are candidates for OCS as well. For OCS we note that by now experimental data for the orientation-dependent ionization
yields exist for both linearly polarized probe pulses (present work), and for circularly polarized probe pulses \cite{Holmegaard10,Dimitrovski:PRA:2011}. The two sets of results for OCS will stimulate further development of theory. In this connection, it is interesting that very recent time-dependent density-functional theory (TDDFT) calculations on OCS in short, linearly polarized laser pulses performed at approximately twice the intensity considered here~\cite{Bandrauk:PRA:2011}, also do not agree with the present experimental trends. In \cite{Bandrauk:PRA:2011} the calculated yield is smaller at $\Theta=90^\circ$ than at $\Theta=0^\circ$ and is larger, again, at $\Theta = 45^\circ$ in agreement with the theoretical model predictions shown in figure 2, but in disagreement with the experimental findings.

\subsection{Benzonitrile, BN}

\begin{figure}
    \centering
\includegraphics[width=0.7\textwidth]{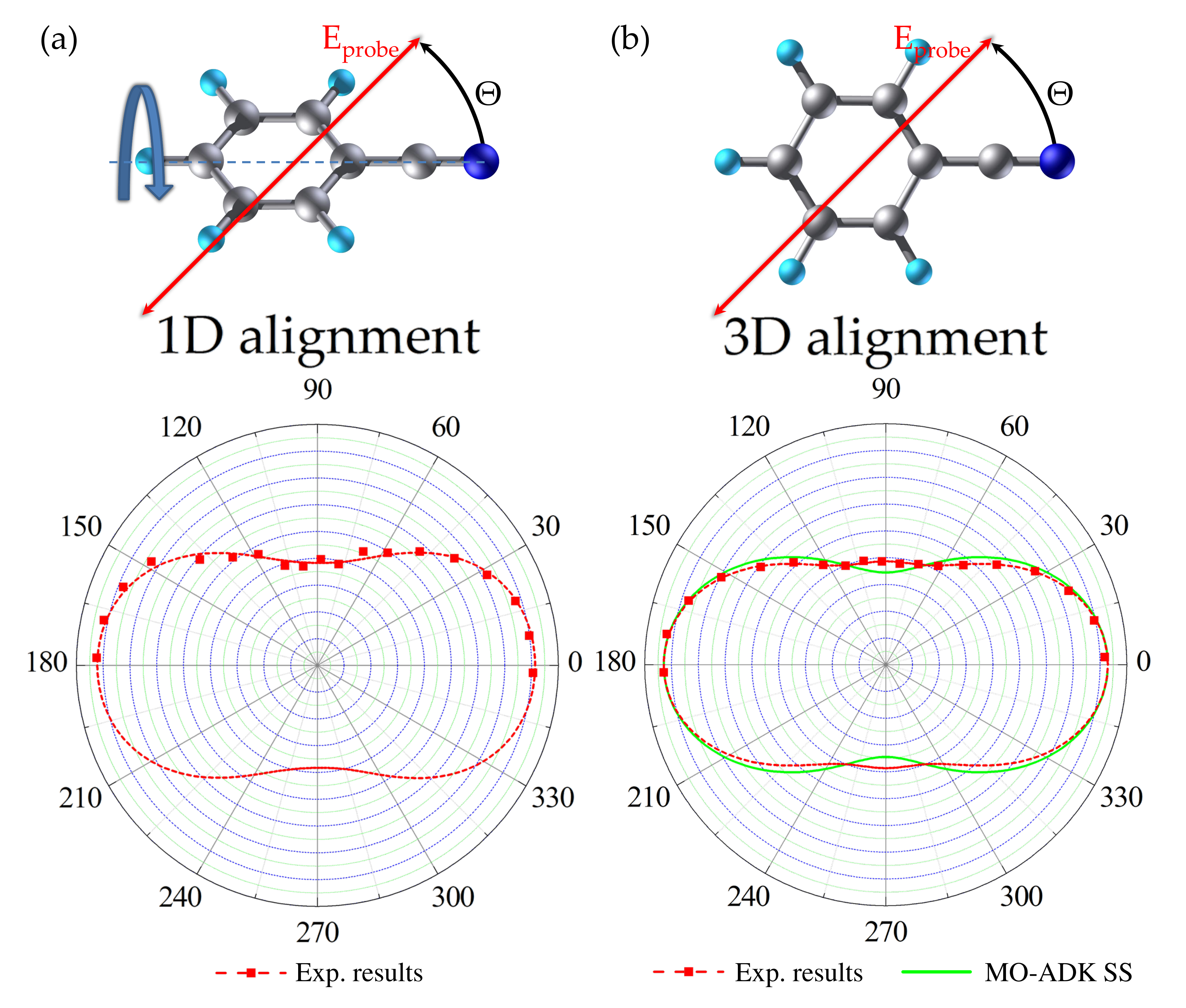}
\caption{
Total ionization yield of (a) 1D and (b) 3D adiabatically aligned BN molecules (red squares, \cost~$=0.89$) as a function of the angle between the major polarization axis of the alignment and the probe pulse. The filled symbols represent the measurements and the dashed curves, to guide the eye, are Legendre polynomial fits. For the case of 3D aligned molecules the theoretically calculated ionization yields (green curve) for confined BN molecules (\cost~$=0.89$) is included (see text). The intensities of the laser pulses used for the experiments and theoretical calculations are $\mathrm{I}_{\mathrm{YAG, 1D}}=\mathrm{I}_{\mathrm{YAG, 3D}}=7.7\times10^{11}$~Wcm$^{-2}$ (10 ns, 1064 nm) and $\mathrm{I}_{\mathrm{probe}}~=~1.2\times10^{14}$~Wcm$^{-2}$.

}
    \label{BN_1D_3D}
\end{figure}

In the case of BN we are able to record the orientation-dependent ionization yields from 1D
and 3D oriented molecules.  First, the orientational dependence of the ionization yield is studied for
1D adiabatically aligned molecules. As for OCS the 1D alignment is induced by a linearly polarized
YAG pulse resulting in confinement of the C-CN axis (the most polarizable axis, see figure 3) along the YAG
polarization with \cost $=0.89$, $\theta_{\mathrm{2d}}$ being the angle between the YAG polarization
and the C-CN axis.  The experimental results, obtained from the total electron signal as a function
of $\Theta$ is displayed in figure \ref{BN_1D_3D}(a). As opposed to OCS, a clear maximum 
in the ionization yield is observed when the polarizations of the YAG
and probe pulses are parallel (probe pulse polarized along the C-CN axis), and a clear minimum
is observed when they are orthogonal.

Next, the orientational dependence of the ion yield is measured for 3D aligned
molecules. BN is an asymmetric top molecule and the linearly polarized YAG pulse leaves
the molecule free to rotate about the arrested C-CN axis. If an elliptically polarized YAG pulse is
employed (here with an intensity ellipticity of 3:1, the value referring to the ratio of the intensities
measured along the major and the minor polarization axis) the C-CN axis remains restricted along the major polarization axis while
the molecular plane aligns with the polarization plane, i.e. the molecule is 3D aligned
\cite{larsen:00, nevo_laser-induced_2009,Hansen:PRA:2011}. The experimental results are shown in
figure \ref{BN_1D_3D}(b).  As for 1D aligned molecules a clear enhancement of the signal is observed
at $\Theta=0$\degree, whereas the yield is almost constant when 60\degree $< \Theta <$ 120\degree.

Compared to OCS, BN is a larger molecule and has a larger dipole moment and larger polarizibilities, see \cite{Hansen:PRA:2011}.
The ionization potential of the HOMO is $I_p=9.79 $ eV  and at the intensity of the experiment the Keldysh parameter
\cite{keldysh:65}  $\gamma \simeq 0.82$. Using
Stark-shifted MO-ADK theory \cite{Holmegaard10,Dimitrovski:PRA:2010,Hansen:PRA:2011,Dimitrovski:PRA:2011}, it is possible to 
explain the orientation-dependent ionization
yields for linearly polarized pulses. This theory models the HOMO of BN with the simplest orbital
that has a nodal plane in the molecular plane of BN, the $2p_x$ orbital (the $x$-axis points toward the
reader in figure \ref{BN_1D_3D}(b)) and includes the Stark shift of the ionization potential
\cite{Hansen:PRA:2011}. These results are then convoluted with the alignment distribution, assuming the most polarizable axis is restricted to deviate from its ideal position only in the directions of the confined molecular plane. This procedure is followed for naphthalene as well. The agreement between the theory and experiment is very good. 

\subsection{Naphthalene, NPTH}

Naphthalene is also an asymmetric top molecule and so the orientational
dependence of the ionization yield can be studied for both 1D and 3D adiabatically aligned
molecules, using linearly and elliptically polarized YAG pulses, respectively. Similar to BN, NPTH has
very large polarizabilities, but zero dipole moment. We note that ionization of NPTH has previously been studied in Ref. \cite{markevitch:03, markevitch:PRA:04}, where the experiment focused on the role of dissociative ionization channels and their signatures in the energy distributions, using longer laser pulses. Here we focus on the orientation-dependent ionization yields.

\begin{figure}
    \centering
\includegraphics[width=0.7\textwidth]{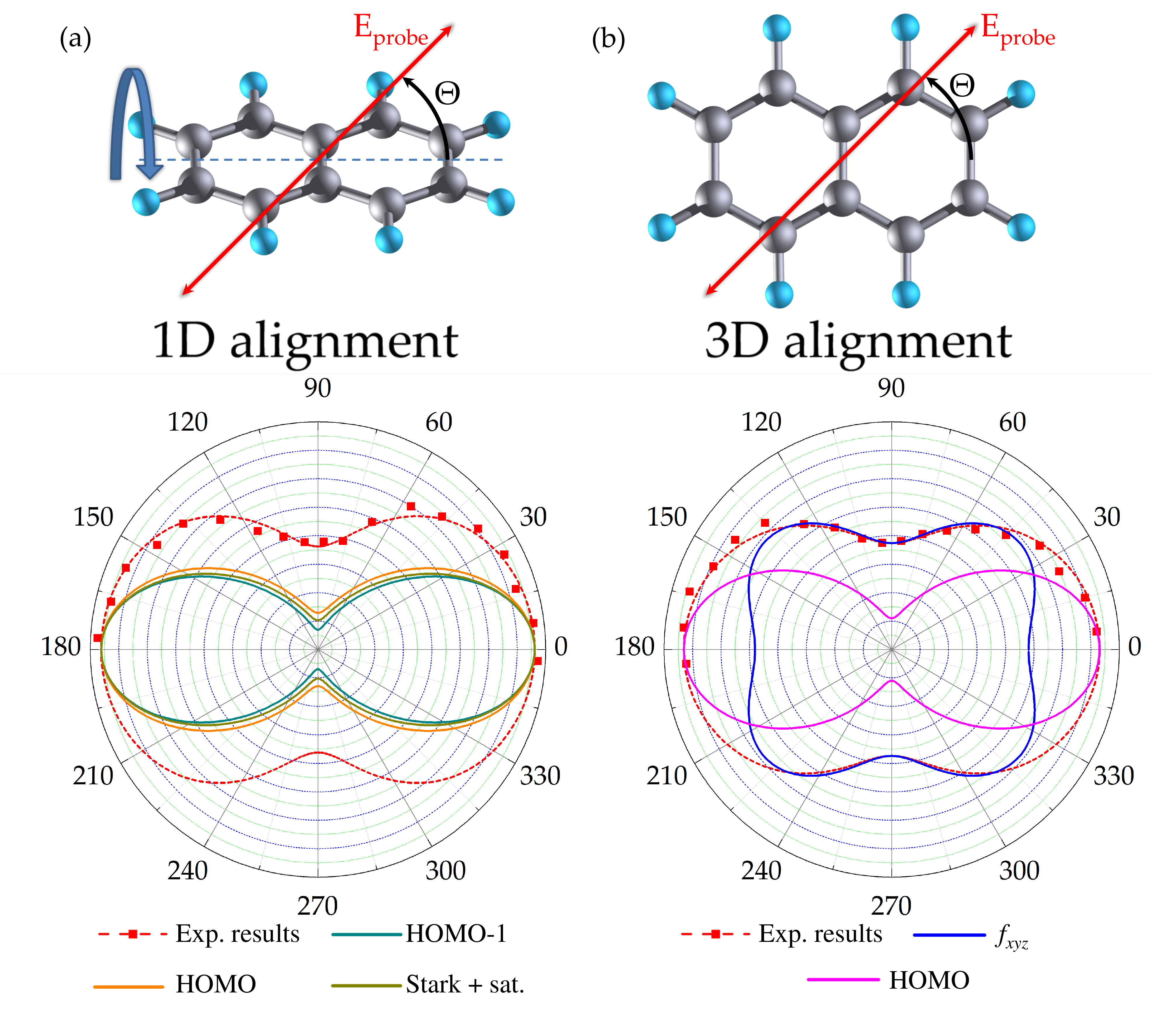}
\caption{
Total ionization yield of (a) 1D and (b) 3D adiabatically aligned NPTH molecules (red squares) as a function of the angle between the major polarization axis of the alignment and the probe pulse. The filled symbols represent the measurements and the dashed curves, to guide the eye, are Legendre polynomial fits.
For the case of 1D aligned molecules the theoretically calculated ionization yields has been modeled using the HOMO (solid orange line), the HOMO-1 (solid cyan line) and the HOMO including stark shifts (solid dark yellow line). The ionization yields from 3D aligned molecules is modeled theoretically using an $f_{xyz}$ orbital (blue line), and the true HOMO (magenta line). All calculations assume \cost~$=0.85$.
The intensities of the laser pulses used for the experiments and theoretical calculations are $\mathrm{I}_{\mathrm{YAG,1D}}=\mathrm{I}_{\mathrm{YAG, 3D}}=7.7\times10^{11}$~Wcm$^{-2}$ (10 ns, 1064 nm) and $\mathrm{I}_{\mathrm{probe}}~=~8.2\times10^{13}$~Wcm$^{-2}$.

}
    \label{Naph_1D_3D}
\end{figure}

For NPTH the only experimental information about the molecular alignment stems from H$^+$ ions created in the Coulomb explosion. In the presence of the YAG pulse, linearly polarized perpendicular to the detector plane, the H$^+$ ion image has the shape of a doughnut. This shows that the long axis of the molecule is confined along the YAG polarization and that the molecular plane is randomly oriented around this axis \cite{Viftrup_3D_2009}. This is in accord with the expectations since the long molecular axis possess the largest polarizability and should, therefore, be the one that is confined along the YAG polarization. The H$^+$ ion images, do, however, not immediately provide quantitative information about the degree of alignment. Due to the high polarizability (anisotropy) of NPTH the alignment is expected to be high \cite{Hansen:PRL:2011}, and of the same order as for BN. Nevertheless we use a conservative estimate for the degree of alignment, \cost~$=0.85$, in the theoretical calculations.
The experimental results of the ionization yield, obtained from the total electron signal as a function of $\Theta$ is displayed in figure \ref{Naph_1D_3D}(a).

If the YAG polarization is changed from linear to elliptical (again with an ellipticity ratio of 3:1) 3D alignment is induced with the long molecular axis along the major component of the elliptically polarized field and the short molecular axis (still in the molecular plane) along the minor axis. The experimental results are shown in
figure \ref{Naph_1D_3D}(b). Comparing figures \ref{Naph_1D_3D}(a) and (b) it can be seen that the measured ionization yields for
1D and 3D aligned molecules show similar features. In particular, the ionization yield peaks
when the polarizations of the alignment and probe pulses are parallel, and reaches a minimum when they are orthogonal. For 1D aligned molecules the modulation ratio, $R=2.1$, is slightly larger than for 3D aligned molecules, $R=2.0$.

To confirm that the angular dependence of the experimental ionization yields displayed in figure 4 indeed results from an orientation dependence rather than from some combined effects of the two laser fields  we repeated the measurements for identical laser intensities but for a backing pressure of the He carrier gas in the supersonic expansion of only 15 bar. When the He pressure is lowered from 90 to 15 bar the rotational temperature increases and the degree of alignment deteriorates significantly \cite{kumarappan_role_2006}. We observed that the pronounced angular dependence of the ion yield, apparent in figure \ref{Naph_1D_3D}
 essentially vanishes. For the measurements on 3D aligned molecules $R$ dropped from 2.0 to 1.3, which is close to
the ratio observed  for purely collisionally aligned molecules~\cite{weida_collisional_1994, pirani_orientation_2001}, that is, with only the probe pulse present, $R=1.1$.

We have performed quantum chemistry calculations~\cite{gamess} of NPTH at the Hartree-Fock level of theory using a TZV basis, including a few diffuse orbitals. 
We obtained the following ionization potentials of interest. $I_p(HOMO) = 8.05$ eV, $I_p(HOMO-1) = 8.84$ eV. The other orbitals are more tightly bound, for example,  $I_p(HOMO-2) = 10.61$ eV. 
The HOMO of NPTH possess three nodal
planes, one coincides with the molecular plane and the other two are orthogonal to each other and
also orthogonal to the molecular plane. The HOMO-1 also has a nodal plane in the molecular plane.
The small energy difference between the HOMO and the HOMO-1 means that they are both expected to contribute to the ionization signal calculated 
by MO-ADK theory at 
experimental Keldysh parameters~\cite{keldysh:65} of a typical value of  $\gamma \simeq 0.9$.
From figure \ref{Naph_1D_3D} we see, as in the case of BN, that the theory reproduces the overall trend of the orientation-dependent
ionization yields. Specifically, in figure \ref{Naph_1D_3D}(a) we consider the case of 1D aligned molecules. The theoretical calculations corresponding to this case were obtained by averaging over the set of results obtained for different angles of orbital rotation around the molecular axis (angles with steps of $10\degree$ are considered), and then convoluting with the alignment distribution. 
Because of the small difference of the ionization potential of HOMO and HOMO-1 for NPTH, and due to the symmetry of the HOMO-1, the angle-dependent ionization rates from both HOMO and HOMO-1 are comparable if not equal, see figure \ref{Naph_1D_3D}(a).
Also, because of the large polarizabilities of NPTH and its cation, we have included the Stark shift of the ionization potential. Including the Stark-shifts does not change much in the orientation-dependent ionization yields, and on the scale of the figure it is difficult to distinguish the three theory
curves. All three curves reproduce the overall trend in the experiment, but not the correct ratio $R$ between the experimental ionization yield along and perpendicular to the molecular axis. As is clear from the figure, the theory predicts a larger $R$ than  the experiment.
This may be due to influence from electronically excited states in the ionization process or possibly to distortion of the HOMO and HOMO-1 by the strong ionizing field.
The same discrepancy between theory and experiment persists for 3D-oriented targets as
well. In figure \ref{Naph_1D_3D}(b), for the 3D-oriented target, we show the theoretical curves from Stark-shifted MO-ADK theory for a HOMO modeled by a $f_{xyz}$ orbital as in \cite{Hansen:PRL:2011}, and by taking the true HOMO of NPTH from our quantum chemistry calculation. In both cases the Stark shifts are included. The curve corresponding to the true HOMO of NPTH, is very similar to the theoretical curves for the 1D-oriented target. On the other hand, in the calculation using the model $f_{xyz}$ orbital, two nodal planes  are visible in the ionization yield. Interestingly, this curve seems to fit very well to the experiment for ionization
perpendicular to the molecular axis, pointing to a possible orbital modification by the field, especially for the geometry with perpendicular orientation of the molecular axis and the field.

\section{Conclusion}

We have presented a set of experimental data for the orientation-dependent strong-field ionization yields for
aligned OCS, BN and NPTH (one linear and two asymmetric top molecules) using linearly polarized probe pulses. 
The orientation-dependent yields were compared with 
simple theoretical models. 
 For OCS the maximum (minimum) ionization yield occurs when the polarization of the ionizing field is perpendicular (parallel) to the internuclear axis. 
 Our theory, as well as numerical TDDFT approaches \cite{Bandrauk:PRA:2011}, predict a maximum elsewhere - in our case at an angle of 45 degrees between the OCS axis and the polarization of field.  We discussed the possibility of the discrepancy being caused by 
the influence of electronically excited states in the multiphoton ionization process. 
It is an open question whether this is correct (indicating a shortcoming of the TDDFT to account for excited states), or whether the discrepancy is due to some other dynamics not accounted for in approaches used so far.

For BN and NPTH the ionization yield is maximized (minimized) when the ionizing field is parallel (perpendicular) to the molecular axis with the largest polarizability. In both cases the present modelling accounted for that behavior. In the case of BN a very good agreement between experimental data and theory was obtained. In the case of NPTH the theory over-estimates the ratio between the ionization yields parallel and perpindicular to the most polarizable axis.

Our results suggest that the angular behaviour of the total ionization yields in linearly polarized fields cannot, in general, be described  by  ionization rates only. In this sense, our analysis suggests that to map out the instantaneous ionization rate of the molecule it is better to use circularly polarized laser pulses because both rescattering with the parent ion and the influence of excited molecular states is minimized.

\section*{Acknowledgements}
The work was supported by the Danish National Research Council (Grant number 10-085430), The Lundbeck
Foundation and the Carlsberg Foundation and the Danish Council for Independent Research (Natural
Sciences).


\section*{References}


\end{document}